\documentclass[%
 reprint,
 amsmath,amssymb,
 aps,
]{revtex4-2}
\usepackage{balance}
\usepackage{xcolor}
\usepackage{graphicx}
\usepackage{hyperref}
\usepackage{tabularx}
\usepackage{amsmath}
\usepackage{svg}
\usepackage{capt-of}
\usepackage{amssymb}
\usepackage[dvipsnames]{xcolor}

\usepackage{dcolumn}
\usepackage{bm}

\begin{document}

\title{Noise-Driven Differentiation via Gene Frustration and Epigenetic Fixation}

\author{Davey Plugers}
\email{davey.plugers@nbi.ku.dk}
\affiliation{Niels Bohr Institute, University of Copenhagen, Blegdamsvej 17, 2100 Copenhagen, Denmark}%

\author{Kunihiko Kaneko}
\email{kunihiko.kaneko@nbi.ku.dk}
\affiliation{Niels Bohr Institute, University of Copenhagen, Blegdamsvej 17, 2100 Copenhagen, Denmark}

\date{\today}

\begin{abstract}
Gene expression in cells is stochastic, yet differentiation can display reproducible timing and stable fate commitment. We develop an analytical theory for a previously identified mechanism in which weakly stable intermediate, or frustrated, gene-expression states are perturbed by stochastic fluctuations and subsequently fixed by slow epigenetic feedback. By eliminating the fast expression dynamics, we show that the differentiation of the slow epigenetic variable is driven by the noise of gene-expression, which can be amplified by regulatory interactions. We derive the logarithmic dependence of onset time for differentiation upon the effective noise intensity, and the input-dependent probability of reaching either fate. We further construct a Waddington-inspired time-dependent probability landscape that visualizes population branching and progressive fate fixation.
\end{abstract}

\maketitle

In biological systems, particularly within cells, stochasticity is unavoidable. As the number of molecules is not necessarily large and transcription often occurs in bursts, biochemical reactions and gene expression dynamics involve substantial stochasticity \cite{blake2003noise,raser2005noise,pedraza2008effects,bax2025gene,raj2008nature,elowitz2002stochastic,mcadams1997stochastic,vinuelas2013quantifying,kaern2005stochasticity,furusawa2005ubiquity,ham2020extrinsic,lin2021disentangling,choi2025protein}. Robustness to such noise is therefore expected to have evolved to enable cells to function properly. In fact, the noise tolerance of gene expression, responses to external signals, spatial patterning, and cell differentiation have been extensively investigated \cite{cortijo2020does,little2013precise,mirabet2012noise,lim2025toward,maini2012turing,suel2007tunability,tkavcik2008role}. 

The positive roles of noise in cellular function have also been discussed \cite{tsimring2014noise,eldar2010functional}, including noise- or discreteness-induced phase transitions \cite{togashi2001transitions}, stochastic resonance \cite{mcdonnell2009stochastic}, noise amplification \cite{chalancon2012interplay,puzovic2023being}, attractor selection by noise \cite{puzovic2023being,pinho2015phenotype,kashiwagi2006adaptive,furusawa2009chaotic}, boundary sharpening \cite{zhang2012noise}, and cell-state exploration \cite{toppen2025noise,ahrends2014controlling}.

Still, how cell differentiation can be both noise-driven and robust to noise or how this may evolve remains elusive \cite{nie2020noise,long2019emergence,wagner1996does}, with recent works discussing the interplay between evolvability, population genetics, noise heritability and their effects on gene regulation architecture \cite{weinreich2025population,dzib2026gene} . In the classical Waddington landscape, differentiation over developmental time is described as a robust process termed {\sl homeorhesis} \cite{waddington1957strategy}, in which cell states move along branching valleys. These branches correspond to differentiated cell types and their fractions, while the timing of branching is robust to intracellular stochasticity. Initially pluripotent cells can transition into various cell fates as intracellular gene-expression patterns and chromatin states change \cite{niwa2007pluripotency,meissner2010epigenetic,lim2013hematopoietic}. 

Such transitions from one to multiple stable states can be modeled through bifurcation mechanisms that represent the cell-fate decision process \cite{huang2007bifurcation,furusawa2001theory,matsushita2020homeorhesis,rand2021geometry,saez2022dynamical,mojtahedi2016cell,ham2025mapping}. However, gene expression levels within cells are noisy, and pluripotent cells must make reliable decisions near critical points for robust differentiation \cite{moris2016transition,guillemin2021noise,mojtahedi2016cell,huang2020decoding,ahrends2014controlling}. This requires suitable regulatory configurations and interactions to ensure that differentiation proceeds reproducibly \cite{moghe2025optimality}.

In our previous work, we identified a class of gene-regulatory networks in which differentiation proceeds through weakly stable frustrated gene-expression states, noise amplification, and slow epigenetic fixation \cite{plugers2026evolution}. That study unveiled qualitatively the mechanism through numerically evolved regulatory networks but did not provide a general analytical description, in particular, its timing, noise dependence, or response to external bias.

In the present work, we analyze the corresponding slow-fast stochastic cell-fate decision process and derive its principal parameter dependencies. Specifically, we determine the noise transmitted from the fast gene-expression dynamics to the slow epigenetic variable, derive the logarithmic scaling of the onset time of nonlinear differentiation, calculate the bias-dependent probability of reaching either fate, and show how regulatory-network structure modifies fate accessibility through noise amplification. These results are compared with direct numerical simulations.

We introduce gene-expression levels $x_i$ and slow epigenetic variables $\theta_i$ ($i=1,\ldots,M$), following Ref.~\cite{mjolsness1991connectionist} (see also Refs.~\cite{salazar2001phenotypic,kaneko2007evolution}):
\begin{subequations}
\begin{equation}
\label{eq:Main}
\frac{dx_i}{dt}
=
F(y_i)-x_i+\sigma_i\eta_i(t)
\end{equation}
\begin{equation}
\label{eq:Epi}
\frac{d\theta_i}{dt}
=
\nu(x_i-\theta_i)
\end{equation}
\end{subequations}
The total regulatory input is $y_i=\sum_{j=1}^{M}J_{ij}x_j+\theta_i+c_i$, and transcription follows the monotonic sigmoidal response $F(y_i)=\tanh(\beta y_i)$, where $\beta$ is the regulatory gain. Gene expression levels are attracted to on/off states. The additive regulatory input is a minimal approximation commonly used in continuous gene-regulatory models, assuming that leading effects of regulatory signals can be represented by a weighted superposition. Cooperative and combinatorial regulation may instead require a more general nonlinear input function, requiring a separate analysis of the reduced dynamics and noise propagation. The independent Gaussian white noises satisfy
\begin{equation}
\left\langle\eta_i(t)\eta_j(t')\right\rangle
=
\delta_{ij}\delta(t-t')
\end{equation}
Equivalently, the stochastic term may be written as $\sigma_idW_i(t)$. The parameter $\sigma_i$ represents an effective local amplitude of gene-expression fluctuations near the frustrated state. In a microscopic chemical-Langevin description, its magnitude would depend on molecular copy numbers, transcriptional bursting, degradation, and extrinsic fluctuations. Our use of additive Gaussian white noise is therefore a local approximation and more general state-dependent or multiplicative noise can be incorporated, but would modify the effective slow-noise intensity and require a separate analysis \cite{coomer2022noise,gillespie2000chemical}.

The interaction $J_{ij}>0$ represents activation of gene $i$ by gene $j$, whereas $J_{ij}<0$ represents repression, and $c_i$ is an external or constitutive input. The variable $\theta_i$ is a coarse-grained epigenetic memory, such as persistent chromatin accessibility or a regulatory modification reinforced by gene expression \cite{matsushita2020homeorhesis,dodd2007theoretical,sneppen2008ultrasensitive,palacios2025analog}. 

Furthermore, the mathematical role of $\theta_i$ does not require it to be a chromatin modification. It may more generally represent any slowly varying intracellular process that integrates regulatory activity and subsequently influences a fate-determining gene-expression program. Examples could include a persistent transcription-factor state, a slowly turning-over regulatory protein, or an intracellular integrator of signaling history. We use the term epigenetic fixation for the chromatin-associated interpretation, while the reduced dynamics apply more broadly to slow fate-controlling memory.

Equation~(\ref{eq:Epi}) is a minimal first-order representation of this slow memory. This causes a slow timescale dynamics such that the fully expressed state $x_i = +1$ is approached when $\sum_j J_{ij}x_j$ exceeds the threshold $-(\theta_i + c_i)$; otherwise, the unexpressed state $x_i = -1$ is approached. \footnote{There are cases with limit cycles or chaotic attractors depending on $J_{ij}$. Differentiation from such states has already been studied in \cite{plugers2026evolution} and will not be discussed here.}

In our previous work \cite{plugers2026evolution}, Eqs.~(\ref{eq:Main}) and (\ref{eq:Epi}) produced saturated differentiated attractors with $x_i\simeq\theta_i\simeq\pm1$. The evolved networks differentiated through the combined effects of frustrated genes, network-mediated noise amplification, and slow epigenetic fixation.

We call a gene \textit{frustrated} when activating and repressing inputs approximately balance, placing its conditional expression level near the activation threshold, $\overline{x}_i\simeq0$. For fixed $\theta_i$, this fast expression state can remain locally stable, while the coupled epigenetic feedback creates an unstable slow direction. Noise supplies the initial symmetry-breaking displacement, which is subsequently amplified and fixed by the slow feedback.

We first isolate a single frustrated gene because it provides the minimal local mechanism governing stochastic fate selection near the activation threshold. Regulatory interactions do not change this basic slow-fast mechanism for \textit{frustrated} genes, but modify its effective coefficients, particularly the noise transmitted to the epigenetic variable. These network coefficient effects are considered below, while other gene expression dynamics such as \mbox{(quasi-)}periodic limit cycles, chaos are beyond the scope of this work.

We first consider a single frustrated gene with $c_i=0$ and $J_{ij}=0$ for $i\neq j$. At fixed $\theta_i=0$, its conditional fixed point is $x_i^*=0$. The fast linear relaxation rate is $\lambda_i=1-\beta J_{ii}$, so the fast expression subsystem is locally stable when $\lambda_i>0$. This condition refers to the $x_i$ dynamics with $\theta_i$ held fixed and is distinct from the stability of the full coupled $(x_i,\theta_i)$ system.

\begin{figure}
    \centering
    \includegraphics[width=1\linewidth]{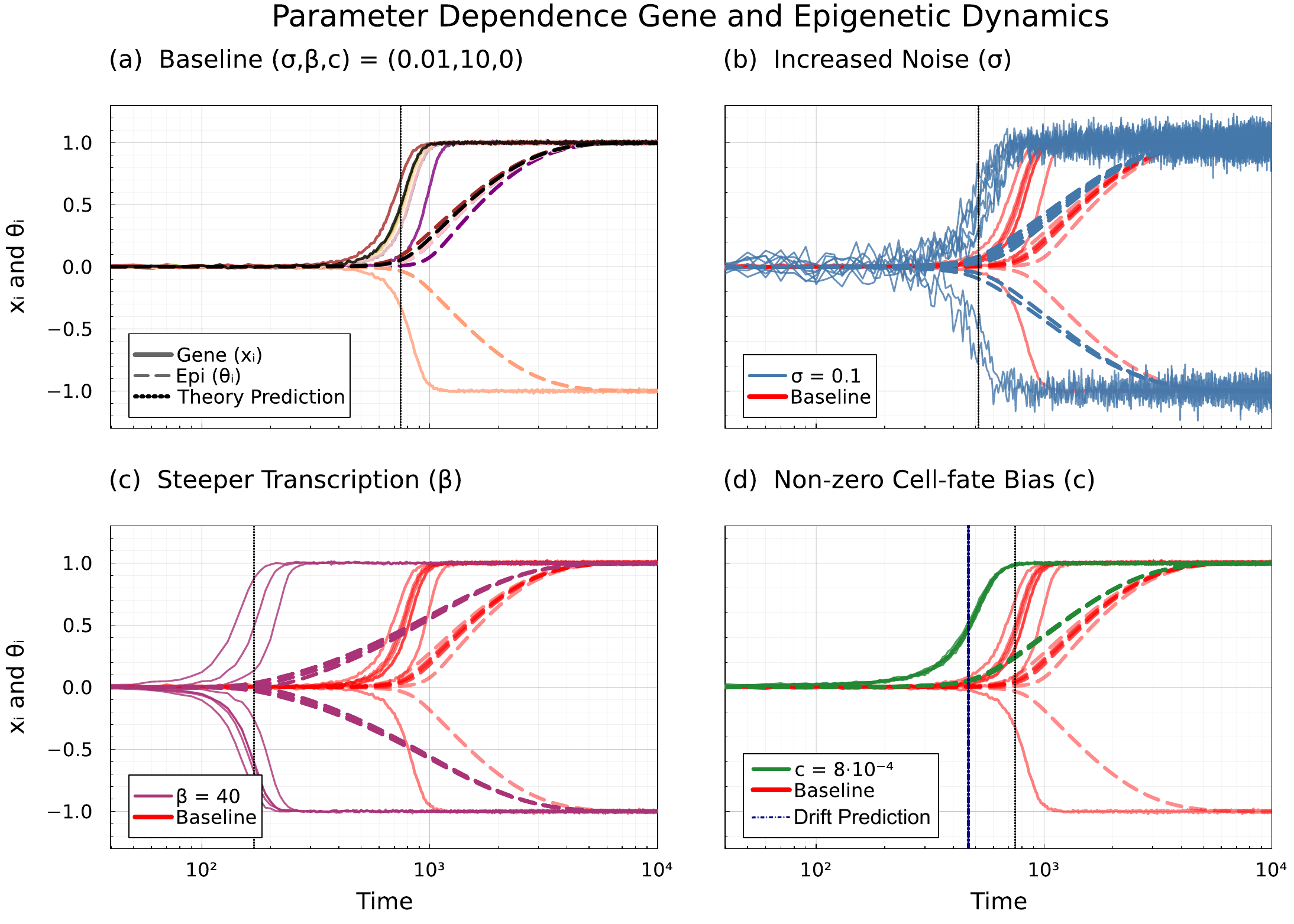}
    \caption{\textbf{(a)} Gene-expression levels $x_i$ (solid) and epigenetic variables $\theta_i$ (dashed) for 8 independent noise realizations. Matching colors denote the same realization. \textbf{(b--d)} The baseline dynamics (red) are compared with changes in the noise amplitude $\sigma_i$ (blue), regulatory gain $\beta$ (purple), and input bias $c_i$ (green). Vertical dotted \textit{Theory Prediction} and dash-dotted \textit{Drift Prediction} lines show the analytical nonlinear-onset estimate $t_c$ and $t_{c,Dr}$ from Eqs.~(\ref{eq:TimeToDiff}) and \ref{eq:DriftTime}. Common parameters are $J_{ii}=0$, $\nu=10^{-3}$, and $\theta_i(0)=0$. Baseline and Panel-specific parameters are indicated in the figure}
    \label{fig:BifurcationSimple}
\end{figure}

Figure~\ref{fig:BifurcationSimple} shows 8 independent realizations of the single-gene model. Differentiation is observed as the initially overlapping trajectories separate toward the positive and negative branches. In our examples, all but one realization have two-sided cell fate commitment at the predicted times. Only the plot for the nonzero bias $c_i$ shows one cell fate and follows a different onset time for differentiation.

Figure~\ref{fig:BifurcationSimple}(a) illustrates the separation of the fast expression variable $x_i$ followed by the slower fixation of the epigenetic variable $\theta_i$. This reflects the central slow--fast mechanism: stochastic expression fluctuations initiate fate selection, while the gradually accumulating epigenetic memory stabilizes the selected state. The vertical dashed lines indicate the analytical estimate of nonlinear onset.

In Fig.~\ref{fig:BifurcationSimple}(b), the frustrated gene fluctuates near $x_i\approx0$ until $T\approx10^{2.5}$, after which the growing epigenetic feedback drives effectively irreversible commitment. Increasing $\sigma_i$ leads to earlier differentiation, but the effect is comparatively weak because the onset time depends only logarithmically on the effective noise.

In Fig.~\ref{fig:BifurcationSimple}(c), increasing the regulatory gain from $\beta=10$ to $40$ produces a stronger reduction in the differentiation time. A larger $\beta$ increases the response of gene expression to changes in $\theta_i$ and therefore strengthens the positive-feedback amplification. In larger regulatory networks, $\beta$ also controls the transmission of upstream fluctuations through the factors $\beta J_{ij}$ in Eq.~(\ref{eq:NoiseAmp}).

Finally, Fig.~\ref{fig:BifurcationSimple}(d) shows that a small input bias $c_i=8\times10^{-4}$ produces unequal fate probabilities. At low noise, the fate opposed by the bias can become sufficiently rare that the sampled population appears effectively monomodal. This does not necessarily imply the loss of a deterministic fixed point, but rather that one fate is improbable to be realized within the finite sample.

The parameters affect the differentiation time through different parts of the reduced dynamics. Increasing $\sigma_i$ increases the initial fluctuation amplitude, but changes the nonlinear-onset time only logarithmically. By contrast, changing $\beta$, $J_{ii}$, or $\nu$ also modifies the unstable slow rate $\mu_i$ and can therefore produce a stronger change in the differentiation time. A nonzero $c_i$ primarily changes the relative accessibility of the two final fates unless the drift induced timescale $t_{c,Dr}$ dominates over the stochastic differentiation time $t_c$ as discussed later.

Now we derive the reduced dynamics for slow $\theta_i$ by eliminating the fast $x_i$ variables. Because the epigenetic rate $\nu$ is small compared with the fast relaxation rate $\lambda_i=1-\beta J_{ii}$, the variable $\theta_i$ changes only weakly during the relaxation of $x_i$. We may therefore treat $\theta_i$ as fixed when determining the conditional quasi-steady expression state and set the deterministic part of $dx_i/dt$ to zero. Linearizing the transcription function near the frustrated state we obtain the fixed point $x_i^*$ for given $\theta_i$:
\begin{equation}
\begin{aligned}
  &x_i^* = \frac{\beta}{1-\beta J_{ii}}\left[\sum_{j\neq i} J_{ij}x_j^* + c_i + \theta_i \right]\\
  &\operatorname{Var}(x_i) = \frac{\sigma_i^2}{2\lambda_i}
\end{aligned}
\end{equation}
Noise then causes fluctuations around the fixed point expression $x_i^*$, given by the variance of the Ornstein-Uhlenbeck process of the perturbations to their mean. This would lead to an effective fast noise magnitude $\sigma_{i,\text{eff}} = \sigma_i/\sqrt{\lambda_i}$ for fluctuations around $x_i^*$.

Following the above argument, the fast gene expression dynamics can be adiabatically eliminated \cite{haken2004introduction}, treating $x_i$ to always be on its fixed point. The noise term in $x_i$ is then transferred to the noise experienced by $\theta_i$ \cite{kaneko1981adiabatic,risken1989fokker}, leading to the Langevin equation of the slow components. Separating the deterministic conditional mean from the fast fluctuations and integrating their autocorrelation gives:
\begin{equation}
\label{eq:CubicTheta}
d\theta_i
=
\left[
\mu_i(\theta_i+\widetilde{c}_i)
-
\kappa_i(\theta_i+c_i)^3
\right]dt
+
\widetilde{\sigma}_idW_i(t)
\end{equation}
where we have defined rescaled parameters:
\begin{equation}
\label{eq:RescaledParams}
\begin{aligned}
\mu_i
&=
\frac{\nu(\beta-1+\beta J_{ii})}{\lambda_i},
\
&\widetilde{c}_i 
&=
\frac{\beta c_i}{\beta-1+\beta J_{ii}},
\\
\kappa_i
&=
\frac{\nu\beta^3}{3\lambda_i^4},
\
&\widetilde{\sigma}_i^2
&=
\frac{\nu^2\sigma_i^2}{\lambda_i^2}
\end{aligned}
\end{equation}

Here, $\mu_i$ is the unstable growth rate of the slow epigenetic mode, $\kappa_i$ controls the nonlinear saturation that stabilizes the two differentiated branches, $\widetilde{c}_i$ and $c_i$ are rescaled and original bias terms that lead to asymmetric cell fate distributions, and $\widetilde{\sigma}_i$ is the noise amplitude transmitted from the fast gene-expression dynamics onto the slow epigenetic variable.

A detailed derivation, as well as a derivation of the noise scaling is given in the Appendix.

Before discussing the analytical estimate of the onset time for differentiation in Eq. \ref{eq:TimeToDiff}, we provide the stability criteria required for a \textit{frustrated} gene to differentiate. The full coupled origin is stable when $\beta(J_{ii}+1)<1$ and becomes unstable along the slow direction when $\beta(1+J_{ii})-1>0$. For $c_i=0$, the cubic approximation has the unstable fixed point $\theta_i=0$ and stable branches:
\begin{equation}
\theta_i^\pm
=
\pm\sqrt{\frac{\mu_i}{\kappa_i}}
\end{equation}
With the full sigmoidal dynamics subsequently saturating near $x_i\simeq\theta_i\simeq\pm1$. 

The fast expression subsystem is stable for fixed $\theta_i$ when $-\lambda_i=\beta J_{ii}-1<0$. Combining both criteria, differentiation therefore occurs in the open interval:
\begin{equation}
J_{ii} \in 
\left]-\frac{\beta-1}{\beta}, \frac{1}{\beta}\right[
\end{equation}
Within this interval, the fast expression state remains conditionally stable while the coupled epigenetic direction is unstable. The endpoints are excluded because $\mu_i=0$ at the lower boundary and $\lambda_i=0$ at the upper boundary.

For the unbiased case\footnote{Because the shifts in the linear and cubic terms differ when $c_i\neq0$, the noise-driven onset-time calculation below is restricted to the unbiased case}, the initially narrow distribution broadens under the unstable linear dynamics until it reaches the scale at which the cubic term becomes relevant. Following the transient-scaling approach of Refs.~\cite{suzuki1976scaling,de1980transient}, the asymptotic nonlinear-onset time is:
\begin{equation}
\label{eq:TimeToDiff}
t_c
\simeq
\frac{1}{2\mu_i}
\ln\left(
\frac{2\mu_i^2}
{\kappa_i\widetilde{\sigma}_i^2}
\right) = \frac{1}{2\mu_i} \ln\left[\frac{6 \mu_i^2(1-\beta J_{ii})^6}{\nu^3 \beta^3 \sigma_i^2}\right]
\end{equation}
The derivation is provided in the Appendix. Equation~(\ref{eq:TimeToDiff}) estimates the onset of nonlinear separation rather than the exact time at which a trajectory reaches its final saturated state. In addition, the differentiation time differs between cells due to stochasticity.

For $c_i\neq0$, differentiation may instead be dominated by deterministic drift. Neglecting noise and using the linearized fast response gives:
\begin{equation}
\label{eq:DriftTime}
t_{c,\mathrm{Dr}}
=
\frac{1}{\nu(\beta_{\mathrm{eff}}-1)}
\ln\left[
\frac{
\beta_{\mathrm{eff}}c_i
\pm(\beta_{\mathrm{eff}}-1)
}{
\beta_{\mathrm{eff}}^2c_i
}
\right]
\end{equation}
where
\begin{equation}
\beta_{\mathrm{eff}}
=
\frac{\beta}{1-\beta J_{ii}}
\end{equation}
The sign is chosen to match the sign of $c_i$. The derivation and the operational threshold used in this estimate are given in the Appendix

\begin{figure}[!h]
    \centering
\includegraphics[width=1\linewidth]{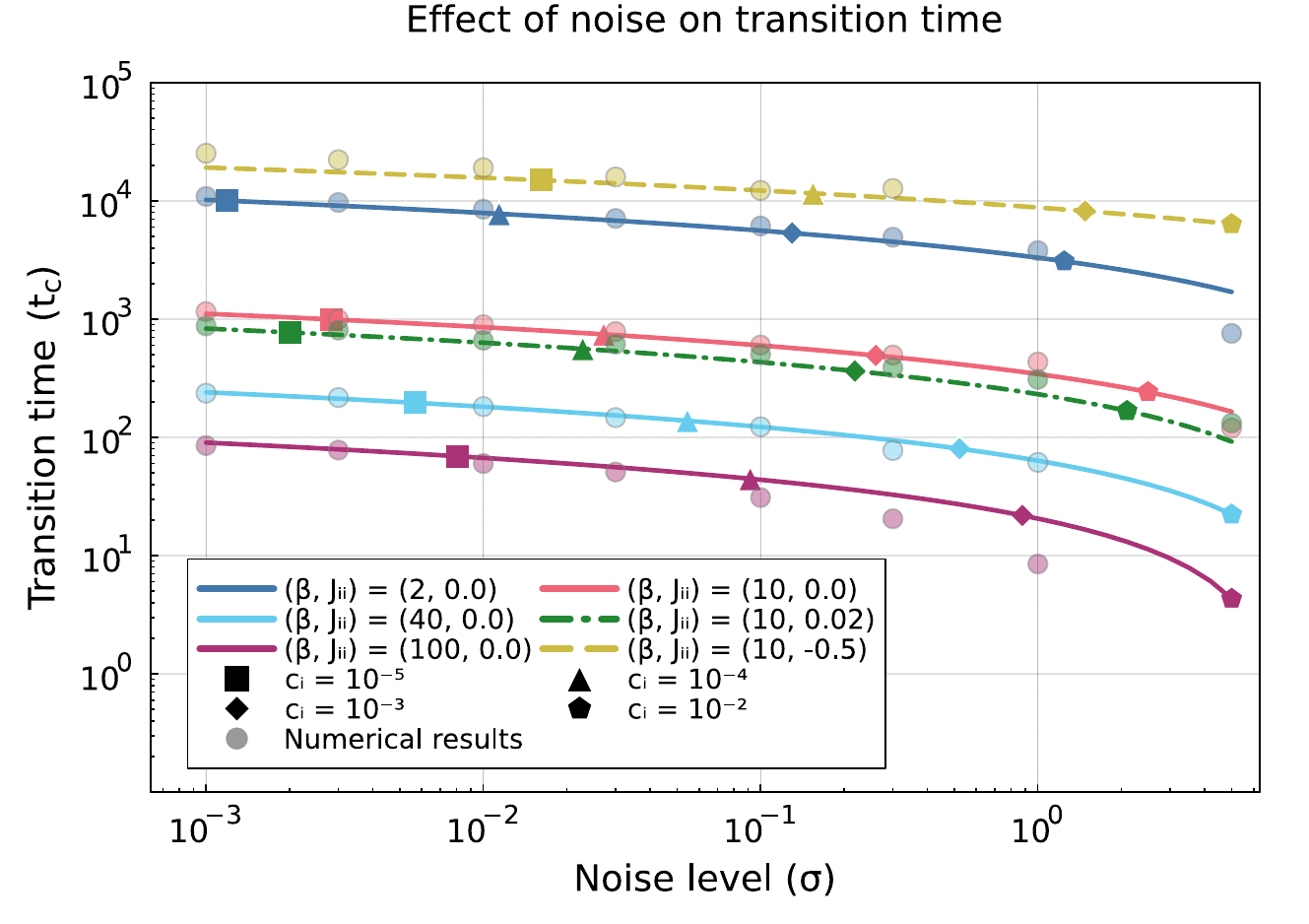}
    \caption{Noise-driven nonlinear-onset time $t_c$ from Eq.~(\ref{eq:TimeToDiff}) as a function of $\sigma_i$ for several values of $\beta$ (lines), compared with the operational commitment time measured in simulations (symbols). Markers indicate where the drift estimate $t_{c,\mathrm{Dr}}$ becomes shorter than the noise-driven estimate. Dashed curves show the effect of nonzero $J_{ii}$ for $\beta=10$. All plotted parameters satisfy $1-\beta J_{ii}>0$, $\nu= 10^{-3}$.}
    \label{fig:TimeToDiff}
\end{figure}

We compare the analytical estimates with numerical simulations in Fig.~\ref{fig:TimeToDiff}. The analytical $t_c$ marks nonlinear onset, whereas the numerical commitment time is defined as the first time after which the absolute moving average of $x_i$ over a window of length $100$ exceeds $0.95$. The two definitions need not coincide exactly, but their parameter dependencies can be compared.

For an isolated frustrated gene, the effective epigenetic noise intensity is:
\begin{equation}
\widetilde{\sigma}_i^2
=
\frac{\nu^2\sigma_i^2}{\lambda_i^2}
\end{equation}
In a hierarchical network, fluctuations transmitted from upstream genes further modify this noise. When feedback and correlations between separate upstream branches are negligible, the effective noise is represented by:
\begin{equation}
\widetilde{\sigma}_i^2
=
\nu^2s_i
\end{equation}
where,
\begin{equation}
\label{eq:NoiseAmp}
s_i
\simeq
\frac{1}{\lambda_i^2}
\left[
\sigma_i^2
+
\sum_{j\neq i}
(\beta J_{ij})^2s_j
\right]
\end{equation}

Thus, both regulatory coupling and weak fast stability amplify the fluctuations experienced by the slow epigenetic dynamics. The corresponding matrix result for a general linearized network is given in the Appendix.

A nonzero input $c_i$ biases the accessibility of the two differentiated branches. As the bias increases relative to the effective noise, the disfavored fate becomes unlikely to be observed in a finite population, even when both deterministic branches remain present. For the linearized biased dynamics, the probability of reaching $\theta_i=+1$ before $\theta_i=-1$, starting from $\theta_i=0$, is:

\begin{align}\label{eq:DifferentiationProb}
    &\mathbb{P}_{0\rightarrow 1}(\mu/\tilde \sigma^2,\tilde c) =\nonumber\\ &\frac{\operatorname{erf}(\sqrt{\mu/\tilde \sigma^2}\left( \tilde c\right)) - \operatorname{erf}(\sqrt{\mu/\tilde \sigma^2}\left(-1 + \tilde c\right))}{\operatorname{erf}(\sqrt{\mu/\tilde \sigma^2}\left(1 + \tilde c\right)) - \operatorname{erf}(\sqrt{\mu/\tilde \sigma^2}\left(-1 + \tilde c\right))}
\end{align}

Here $\mathbb{P}_{0\rightarrow 1}$ is the exact hitting probability for the linearized reduced equation with absorbing thresholds at $\theta_i=\pm1$ and an approximation to the full nonlinear dynamics. The factors $\mu/\tilde \sigma^2$ and $\tilde c$ are rescaled parameters representing the effect of noise and the effect of bias in the system respectively. Dependence of $\mathbb{P}_{0\rightarrow 1}$ on $\sigma$, $\nu$, $\beta$ is plotted in the Appendix.

\begin{figure}[!h]
    \centering
    \includegraphics[width=1\linewidth]{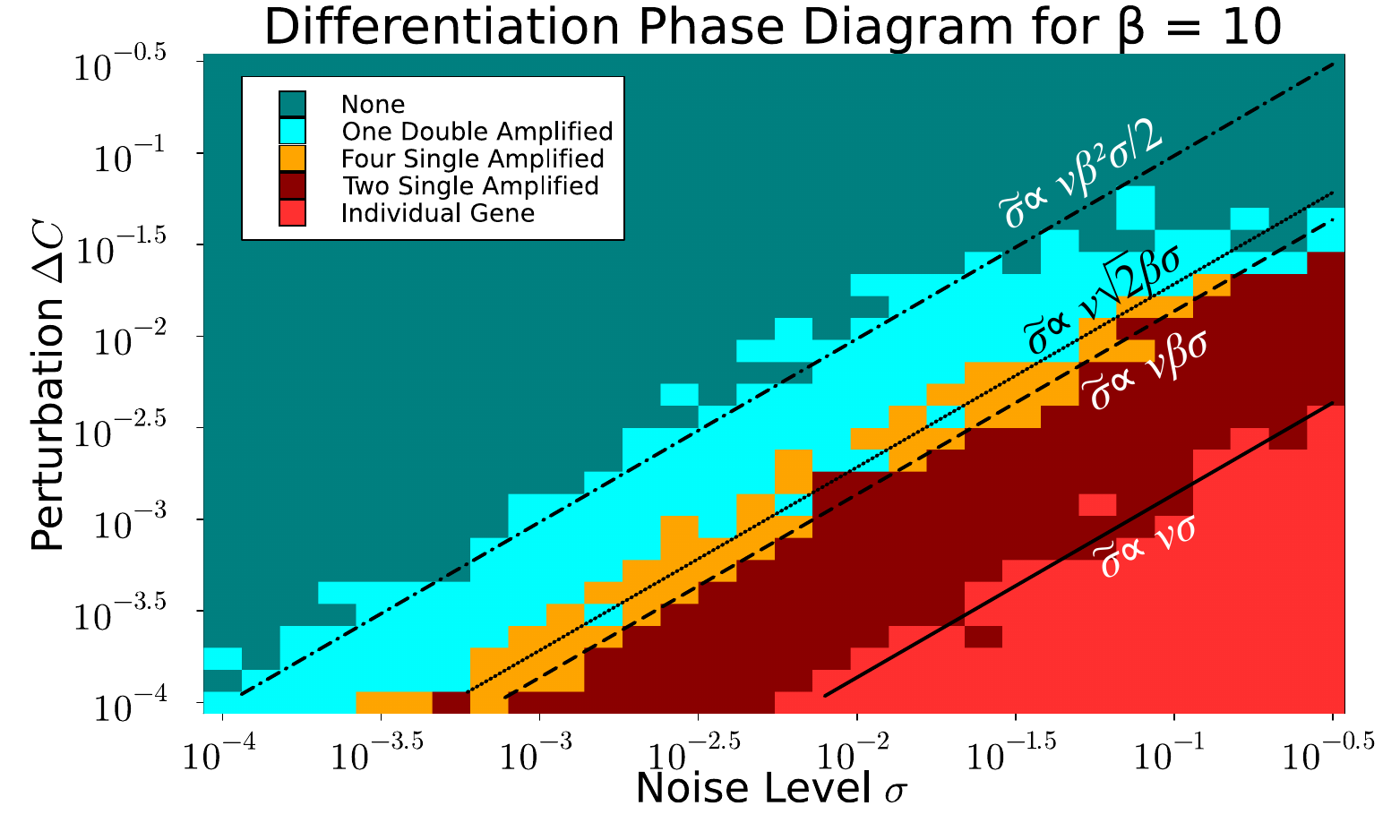}
    \caption{Boundary for observing both differentiated fates in a population of 16 independent cells. Symbols show simulations, and analytical boundaries are obtained from Eqs.~(\ref{eq:NoiseAmp}) and (\ref{eq:DifferentiationProb}) using the topology-dependent effective noise. Color denotes the least noise-amplifying architecture for which both fates are observed: Individual Gene, Two Single-Amplified, Four Single-Amplified, and One Double-Amplified\footnote{Note that since there is no canceling signal for the intermediate gene, we set $c_j = -J_{jk}$ and $c_k = 1000$ such that $y_j = 0$, and $x_k=+1$ ensuring that $x_j$ remains a frustrated gene.}. The analytical criterion is $1-P_{0\rightarrow1}^{16}=1/2$, corresponding that at least one cell adopts the minority fate. Parameters are $\beta=10$, $\nu=10^{-3}$, $J_{ii}=0$, $J_{ij}=\{\textcolor{red}{0}; \textcolor{Maroon}{\pm1} ; \textcolor{orange}{\pm1\&\pm1}, \textcolor{cyan}{+1}\}$, and $J_{jk} = \textcolor{cyan}{+1}$ with colors indicate the respective networks.}
    \label{fig:phasediagram}
\end{figure}
As shown in figure \ref{fig:phasediagram}, this analytical estimate agrees well with the numerical results. To illustrate how network structure modifies fate accessibility, we compare four regulatory architectures: an isolated output gene, an output receiving two independent inputs, an output receiving four independent inputs, and a two-step pathway in which fluctuations are transmitted through an upstream frustrated gene \footnote{For a saturated intermediate gene, fluctuations transmitted through its regulatory inputs are strongly attenuated because the local susceptibility $F'(y_i) = \beta \text{sech}^2(\beta y_i)$ is small. Its own additive expression noise is not eliminated, but network-mediated amplification through that node is greatly reduced due to such saturation.}. These architectures share the same local differentiation mechanism but differ in the noise experienced by the output gene. Figure~\ref{fig:phasediagram} shows the bias $\Delta c$ at which the minority fate ceases to be observed reliably. The analytical boundary is defined by $P_{0\rightarrow1}=\sqrt[16]{1/2}$, so that there is a $50\%$ probability that at least one of 16 cells adopts the minority fate.  

\begin{figure}[!h]
    \centering
    \includegraphics[width=1\linewidth]{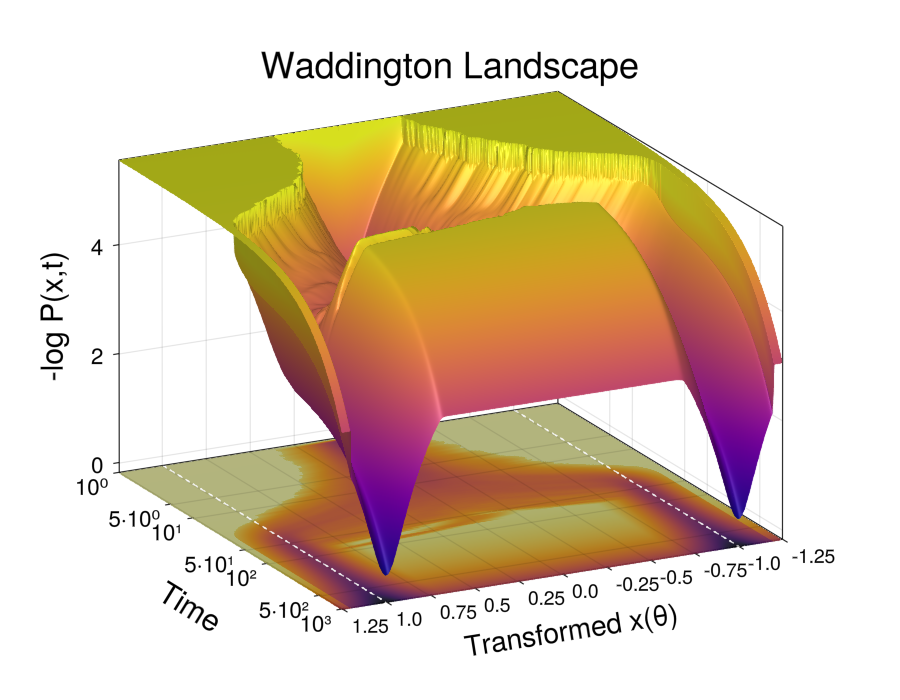}
    \caption{Waddington-inspired time-dependent probability landscape for $(\beta,J_{ii},\sigma_i)=(10,0.01,0.05)$. The quantity $U(x,t)=-\ln P(x,t)$ changes from an initially unimodal valley near $x=0$ to two valleys near the differentiated branches $x\simeq\pm1$. The white dashed line indicates the two bimodal branches $x_i = \pm 1$ to which the system converges. A linear visual tilt has been added to the displayed surface to indicate the direction of developmental time and is not part of $U(x,t)$. Values of the landscape are projected onto the xy plane for clarity.}
    \label{fig:waddington}
\end{figure}

Finally, using Eq.~(\ref{eq:CubicTheta}), we calculate the evolving distribution of $\theta_i$ and construct the Waddington-inspired time-dependent probability landscape shown in Fig.~\ref{fig:waddington}. We define its height as $U(x,t)=-\ln P(x,t)$ after mapping the reduced epigenetic distribution back to the expression variable. We do this by defining the corresponding fixed point in function of the epigenetic variable ($x_i^*(\theta_i)$). This object represents the evolving population distribution and should not be interpreted as a time-independent potential generating the dynamics \footnote{Waddington landscapes have also been constructed from dynamical flows \cite{wang2011quantifying,bhattacharya2011deterministic,boukacem2024waddington}, whereas time-dependent branching associated with slow epigenetic dynamics is emphasized in Refs.~\cite{matsushita2020homeorhesis,rand2021geometry,saez2022dynamical}.}

We developed an analytical description of differentiation from a frustrated gene-expression state through stochastic symmetry breaking and slow epigenetic fixation. Noise initiates the displacement from the frustrated state, while the subsequent amplification is controlled mainly by the unstable slow rate $\mu_i$. The resulting onset time depends only logarithmically on the effective noise, allowing differentiation to require noise while retaining comparatively reproducible timing.

An input bias changes the accessibility of the two differentiated branches, while regulatory interactions shift this accessibility by amplifying noise. These results connect slow-fast stability, differentiation timing, fate probability, and network-mediated noise amplification within a common reduced description.

Although we illustrated differentiation into two fates using a single gene, the present study provides a basis for differentiation driven by noise and epigenetic fixation in more general and complex gene regulatory networks. The present model offers a minimal framework and can readily be extended to differentiation into multiple cell types. Incorporating spatial patterning and signal-induced differentiation into this framework may provide these processes with robustness to noise through $\theta$, which will be explored in future work. In this sense, homeorhesis is expected to emerge more generally in such systems.

The role of noise in generating multiple cell fates has been discussed and observed experimentally in a wide range of biological systems \cite{eldar2010functional,balazsi2011cellular,guillemin2021noise,bax2025gene}. One example is variability arising from bet-hedging differentiation strategies, as observed in plants \cite{abley2024bet} and microbes \cite{grimbergen2015microbial,liu2016use,urban2018buffering}. Experimental reduction of noise has also been linked to a loss of cell fates \cite{maamar2007noise}, further supporting the role of noise in differentiation.

Finally, the present model is consistent with experiments showing that stem cells can exist near critical transition states \cite{bargaje2017cell}, that cell-to-cell variability increases during developmental transitions \cite{gao2023single,canham2010functional}, and that frustrated or intermediate regulatory states can arise under competing signals \cite{yao2025role}. It is also related to theoretical studies of stochastic switching \cite{wells2015control}, regulatory-network configurations that control effective noise \cite{bojer2022robust}, and noise-induced transitions generated by slow-fast positive feedback \cite{zhang2007linking}.

In this interpretation, $\theta_i$ represents a slowly varying, fate-controlling memory that integrates regulatory activity over time and feeds back onto gene-expression dynamics. One possible realization is a persistent chromatin state involving DNA methylation, histone modifications, or chromatin accessibility. More specifically, targeted chromatin-editing experiments have demonstrated analog epigenetic memory, in which distinct grades of DNA methylation were associated with corresponding persistent gene-expression levels rather than only binary expressed and silenced states \cite{palacios2025analog}. These expression states remained after the initial chromatin perturbation, indicating that chromatin can retain quantitative information about previous regulatory conditions. The accompanying model further suggested that analog memory can arise when DNA methylation is not locked into strong positive feedback with the repressive histone modification H3K9me3, whereas stronger coupling between chromatin marks may favor more binary memory. These findings provide a concrete biological motivation for representing chromatin-associated memory by a continuous variable such as $\theta_i$, whose magnitude can encode different degrees of persistent transcriptional activation or repression. Nevertheless, because these experiments used an engineered reporter system, they support the plausibility of continuous epigenetic memory without establishing that the same molecular kinetics underlie the developmental transitions considered here.

The mathematical role of $\theta_i$ is also not restricted to processes conventionally classified as epigenetic. It may more generally represent any slower intracellular process that integrates regulatory history and subsequently influences cell fate. For example, the time integral of BMP signaling has been shown to be a strong predictor of cell fate \cite{teague2024time}. Although accumulated BMP signaling is not itself an epigenetic modification, it illustrates how an intracellular variable can retain information about past signals and control a downstream fate decision. Chromatin modifications, persistent transcription-factor states, slowly turning-over regulatory proteins, and intracellular signaling integrators may therefore belong to the same effective dynamical class when they evolve more slowly than the gene-expression variables, retain a memory of previous regulatory activity, and feed back onto fate selection. The present model should consequently be interpreted as a minimal description of slow fate-controlling memory, with epigenetic fixation representing one biologically important realization rather than the only possible mechanism.

\begin{acknowledgments}
 We are supported by the Novo Nordisk Foundation Grant No. NNF21OC0065542. We would also like to thank Philip Cherian and Tuan Pham for feedback on the figures. The data that support the findings of this article are openly available \footnote{\href{https://github.com/DaveyPlugers/Noise-Driven-Differentiation-via-Gene-Frustration-and-Epigenetic-Fixation/tree/main}{Public GitHub Repository by Davey Plugers: https://github.com/DaveyPlugers/Noise-Driven-Differentiation-via-Gene-Frustration-and-Epigenetic-Fixation/tree/main}   Figures and scripts to analyze results have been made available. Created on 08/04/2026}\nocite{wang2011quantifying,bhattacharya2011deterministic,boukacem2024waddington}
\nocite{matsushita2020homeorhesis,rand2021geometry,saez2022dynamical}
 \end{acknowledgments}
\bibliography{apssamp} 
\clearpage

\onecolumngrid
\section*{Appendix}
\twocolumngrid

\addcontentsline{toc}{section}{Appendix}

\noindent\textit{Rescaled variable and critical differentiation time.—}
Starting from the reduced dynamics for an isolated frustrated gene,
\begin{equation}
\frac{d\theta_i}{dt}
=
\mu_i(\theta_i+\widetilde c_i)
-
\kappa_i(\theta_i+c_i)^3
+
\widetilde\sigma_i\eta_i(t)
\label{eq:ReducedBiasedEndMatter}
\end{equation}
where
\begin{equation}
\widetilde c_i
=
\frac{\beta c_i}
{\beta-1+\beta J_{ii}}
\end{equation}
We emphasize that $c_i$ and $\widetilde c_i$ represent different shifts.
The quantity $c_i$ appears inside the transcription function and therefore
inside the cubic correction, whereas $\widetilde c_i$ is the corresponding
shift of the linear reduced drift.

Defining
\begin{equation}
y_i=\theta_i+\widetilde c_i
\end{equation}
gives
\begin{equation}
\frac{dy_i}{dt}
=
\mu_i y_i
-
\kappa_i
\left[
y_i+(c_i-\widetilde c_i)
\right]^3
+
\widetilde\sigma_i\eta_i(t)
\label{eq:ShiftedBiasedDynamics}
\end{equation}
Thus, the biased nonlinear equation cannot generally be written as
$\dot y_i=\mu_i y_i-\kappa_i y_i^3+\widetilde\sigma_i\eta_i$ unless we have $\beta\gg1$ and $\beta J_{ii} \approx0$. Performing the derivation with such a bias provides a small correction term on the critical transition time $t_c$, but this correction is negligible in the limit $c_i \Rightarrow0$, whereas for larger $c_i$ we find that the drift induced timescale $t_{c,Dr}$ dominates over $t_c$. Rather than including such a correction for $t_c$, we instead derive these two timescales separately, as noise-induced and drift-induced, and estimate which occurs first.

Thus, here the noise-driven critical differentiation time is derived for the
unbiased case $c_i=0$, for which $\widetilde c_i=0$. Equation
(\ref{eq:ReducedBiasedEndMatter}) then reduces to:
\begin{equation}
\frac{d\theta_i}{dt}
=
\mu_i\theta_i
-
\kappa_i\theta_i^3
+
\widetilde\sigma_i\eta_i(t)
\label{eq:ReducedUnbiasedEndMatter}
\end{equation}
During the initial stage, $\theta_i$ is small and the cubic term may be
neglected:
\begin{equation}
\frac{d\theta_i}{dt}
=
\mu_i\theta_i
+
\widetilde\sigma_i\eta_i(t)
\label{eq:LinearUnbiasedEndMatter}
\end{equation}
We retain the rescaled stochastic variable
\begin{equation}
\Xi_i(t)
=
\theta_i(t)e^{-\mu_i t}
\label{eq:XiEndMatter}
\end{equation}
It satisfies
\begin{equation}
\frac{d\Xi_i}{dt}
=
\widetilde\sigma_i e^{-\mu_i t}\eta_i(t)
\end{equation}
For $\theta_i(0)=0$,
\begin{equation}
\Xi_i(t)
=
\widetilde\sigma_i
\int_0^t e^{-\mu_i s}\eta_i(s)ds
\end{equation}
Using
$\langle\eta_i(s)\eta_i(s')\rangle=\delta(s-s')$,
its variance is
\begin{align}
\left\langle\Xi_i^2(t)\right\rangle
&=
\widetilde\sigma_i^2
\int_0^t e^{-2\mu_i s}ds
\\
&=
\frac{\widetilde\sigma_i^2}{2\mu_i}
\left(1-e^{-2\mu_i t}\right)
\label{eq:XiVarianceEndMatter}
\end{align}
Since $\theta_i=\Xi_i e^{\mu_i t}$, the variance of the original variable is
\begin{equation}
\left\langle\theta_i^2(t)\right\rangle
=
\frac{\widetilde\sigma_i^2}{2\mu_i}
\left(e^{2\mu_i t}-1\right)
\label{eq:ThetaVarianceEndMatter}
\end{equation}
The nonlinear scale is estimated by balancing the magnitudes of the linear
and cubic drift terms,
\begin{equation}
\mu_i|\theta_i|
\sim
\kappa_i|\theta_i|^3
\end{equation}
which gives the critical threshold:
\begin{equation}
\theta_{i,c}^2
=
\frac{\mu_i}{\kappa_i}
\end{equation}
We define the critical differentiation time by
\begin{equation}
\left\langle\theta_i^2(t_c)\right\rangle
=
\frac{\mu_i}{\kappa_i}
\end{equation}
Using Eq.~(\ref{eq:ThetaVarianceEndMatter}), this gives
\begin{equation}
\boxed{
t_c
=
\frac{1}{2\mu_i}
\ln\left(
1+
\frac{2\mu_i^2}
{\kappa_i\widetilde\sigma_i^2}
\right)
}
\label{eq:CriticalTimeFinite}
\end{equation}
When
$2\mu_i^2/(\kappa_i\widetilde\sigma_i^2)\gg1$, the asymptotic result is
\begin{equation}
\boxed{
t_c
\simeq
\frac{1}{2\mu_i}
\ln\left(
\frac{2\mu_i^2}
{\kappa_i\widetilde\sigma_i^2}
\right)
}
\label{eq:CriticalTimeAsymptotic}
\end{equation}
Substituting
\begin{equation}
\kappa_i
=
\frac{\nu\beta^3}
{3(1-\beta J_{ii})^4},
\qquad
\widetilde\sigma_i^2
=
\frac{\nu^2\sigma_i^2}
{(1-\beta J_{ii})^2}
\end{equation}
we obtain
\begin{equation}
\boxed{
t_c
\simeq
\frac{1}{2\mu_i}
\ln\left[
\frac{
6\mu_i^2(1-\beta J_{ii})^6
}{
\nu^3\beta^3\sigma_i^2
}
\right]
}
\label{eq:CriticalTimeParameters}
\end{equation}
For $c_i\neq0$, the initial dynamics contain a deterministic bias
$\mu_i\widetilde c_i$. In that case, differentiation can be drift-dominated,
and we use the separate drift-time estimate derived below rather than adding
a linear correction to the unbiased noise-driven result.\\
\\

\noindent\textit{Drift-induced differentiation time with self-interaction.—}
We next estimate the onset time when differentiation is dominated by the drift generated by $c_i$. We neglect noise and use the linearized conditional response $x_i^*\simeq\beta_{\mathrm{eff}}(\theta_i+c_i)$ until the operational threshold $|\beta_{\mathrm{eff}}(\theta_i+c_i)|=1$ is reached. This gives an approximate drift time rather than the exact transition time of the full sigmoidal dynamics:

\begin{align}
    x_i^*  &\approx \beta_{\rm eff}(\theta_i + c_i) + \mathcal{O}((\theta_i+c_i)^3)\\
    \frac{d\theta_i}{dt} &= \nu(\beta_{\rm eff} c_i + \theta_i (\beta_{\rm eff}-1))\\ 
   \beta_{\rm eff} &= \frac{\beta}{1-\beta J_{ii}}
\end{align}
By integration we find an expression for $\theta_i(t)$:
\begin{equation}
    \theta_i(t) = \frac{\beta_{\rm eff} c_i}{1-\beta_{\rm eff}} \left[ 1 - e^{\nu (\beta_{\rm eff}-1) t} \right]
\end{equation}
The drift-induced differentiation time $t_{c,Dr}$ is then found when $\beta_{\rm eff} (\theta_i + c_i) = \pm 1$:
\begin{equation}
    \theta_i(t_{c,Dr}) = \pm \frac{1}{\beta_{\rm eff}} - c_i = \frac{\beta_{\rm eff} c_i}{1-\beta_{\rm eff}} \left[ 1 - e^{\nu (\beta_{\rm eff}-1) t_{c,Dr}} \right]
\end{equation}
Solving for $t_{c,Dr}$ gives:
\begin{equation}
\boxed{
    t_{c,Dr} = \frac{1}{\nu (\beta_{\rm eff}-1)} \ln \left[ \frac{\beta_{\rm eff} c_i \pm (\beta_{\rm eff}-1)}{\beta_{\rm eff}^2 c_i} \right]
}
\end{equation}

\noindent\textit{Noise amplification and fluctuation response.—}
We separate the fast expression variable into its deterministic conditional mean $\overline{x}_i$ and a stochastic fluctuation $\delta x_i$ where this conditional mean is equal to the fixed point $x_i^*$ of the system. Near such a frustrated state, the linearized fluctuation dynamics are:
\begin{equation}
x_i(t)=\overline{x_i}(t)+\delta x_i(t)
\end{equation}
Near the frustrated state, the linearized fluctuation dynamics are
\begin{equation}
\label{eq:LinearFluctuationDynamics}
d\delta x_i
=
\left[
-\lambda_i\delta x_i
+
\sum_{j\neq i}A_{ij}\delta x_j
\right]dt
+
\sigma_idW_i(t)
\end{equation}
where
\begin{equation}
\lambda_i=1-\beta J_{ii},
\qquad
A_{ij}=\beta J_{ij}
\end{equation}
For an isolated gene, Eq. (\ref{eq:LinearFluctuationDynamics}) reduces to an Ornstein-Uhlenbeck process,
\begin{equation}
d\delta x_i
=
-\lambda_i\delta x_idt
+
\sigma_idW_i(t)
\end{equation}
Its stationary autocorrelation is
\begin{equation}
C_{ii}(\tau)
=
\left\langle
\delta x_i(t)\delta x_i(t+\tau)
\right\rangle
=
\frac{\sigma_i^2}{2\lambda_i}
e^{-\lambda_i|\tau|}
\label{eq:OUAutocorrelationAppendix}
\end{equation}
The equal-time variance of the fast variable is therefore
\begin{equation}
\boxed{
\operatorname{Var}(x_i)
=
C_{ii}(0)
=
\frac{\sigma_i^2}{2\lambda_i}
=
\frac{\sigma_i^2}
{2(1-\beta J_{ii})}
}
\label{eq:FastVariableVarianceAppendix}
\end{equation}
If an effective fast-variable noise amplitude is defined by
\begin{equation}
\operatorname{Var}(x_i)
=
\frac{\sigma_{i,\mathrm{eff}}^2}{2}
\end{equation}
then
\begin{equation}
\boxed{
\sigma_{i,\mathrm{eff}}
=
\frac{\sigma_i}{\sqrt{\lambda_i}}
=
\frac{\sigma_i}
{\sqrt{1-\beta J_{ii}}}
}
\label{eq:FastEffectiveNoiseAmplitude}
\end{equation}
The square root appears because $\sigma_{i,\mathrm{eff}}$ is a noise amplitude, whereas Eq.~(\ref{eq:FastVariableVarianceAppendix}) gives a variance.\\
Applying Itô's rule to $\delta x_i^2$, and then taking the expectation, gives
\begin{equation}
\frac{dC_{ii}}{dt}
=
-2\lambda_i C_{ii}
+
2\sum_{j\neq i}A_{ij}C_{ij}
+
\sigma_i^2
\label{eq:VarianceEvolution}
\end{equation}
where
\begin{equation}
C_{ij}
=
\left\langle
\delta x_i\delta x_j
\right\rangle
\end{equation}
At stationarity, ($dC_{ii}/dt=0$), and hence
\begin{equation}
C_{ii}
=
\frac{\sigma_i^2}{2\lambda_i}
+
\frac{1}{\lambda_i}
\sum_{j\neq i}A_{ij}C_{ij}
\label{eq:StationaryVarianceBalance}
\end{equation}
To estimate the cross-covariance, consider a direct hierarchical connection ($j\rightarrow i$), while neglecting feedback from gene i to j and correlations with other upstream branches. Applying Itô's rule to $\delta x_i\delta x_j$ gives
\begin{equation}
\frac{dC_{ij}}{dt}
=
-(\lambda_i+\lambda_j)C_{ij}
+
A_{ij}C_{jj}
\label{eq:CrossCovarianceEvolution}
\end{equation}
The first term describes the combined relaxation of the two variables, whereas the second describes fluctuations in $x_j$ transmitted to $x_i$. At stationarity,
\begin{equation}
C_{ij}
\simeq
\frac{A_{ij}C_{jj}}
{\lambda_i+\lambda_j}
\label{eq:StationaryCrossCovariance}
\end{equation}
Substitution into Eq.~(\ref{eq:StationaryVarianceBalance}) gives
\begin{equation}
\boxed{
\operatorname{Var}(x_i)
\simeq
\frac{\sigma_i^2}{2\lambda_i}
+
\sum_{j\neq i}
\frac{
A_{ij}^2\operatorname{Var}(x_j)
}{
\lambda_i(\lambda_i+\lambda_j)
}
}
\label{eq:PropagatedFastVariance}
\end{equation}
The slow epigenetic variable integrates the fluctuations of $x_i$ through
\begin{equation}
\frac{d\theta_i}{dt}
=
\nu(x_i-\theta_i)
\end{equation}
For a finite time interval $T$, the variance accumulated from the fluctuating contribution is:
\begin{equation}
\label{eq:FiniteTimeIntegratedVariance}
\operatorname{Var}
\left[
\nu\int_0^T\delta x_i(t)dt
\right]
=
2\nu^2
\int_0^T
(T-\tau)C_{ii}(\tau)d\tau
\end{equation}
For $T\gg\lambda_i^{-1}$, this becomes
\begin{equation}
\operatorname{Var}
\left[
\nu\int_0^T\delta x_i(t)dt
\right]
\simeq
\nu^2T
\int_{-\infty}^{\infty}
C_{ii}(\tau)d\tau
=
\nu^2T
\frac{\sigma_i^2}{\lambda_i^2}
\end{equation}
The factor $2$ in Eq.~(\ref{eq:FiniteTimeIntegratedVariance}) follows from the symmetry $C_{ii}(-\tau)=C_{ii}(\tau)$\\
Matching this long-time variance to an effective reduced equation,
\begin{equation}
d\theta_i
=
\cdots dt
+
\widetilde{\sigma}_idW_i(t)
\end{equation}
gives
\begin{equation}
\boxed{
\tilde{\sigma}_i^2
=
\frac{\nu^2\sigma_i^2}{\lambda_i^2}
=
\frac{\nu^2\sigma_i^2}
{(1-\beta J_{ii})^2}
}
\label{eq:SlowNoiseIntensity}
\end{equation}
or equivalently,
\begin{equation}
\boxed{
\tilde{\sigma}_i
=
\frac{\nu\sigma_i}{\lambda_i}
=
\frac{\nu\sigma_i}
{1-\beta J_{ii}}
}
\label{eq:SlowNoiseAmplitude}
\end{equation}
The two noise scalings therefore describe different quantities:
\begin{equation}
\operatorname{Var}(x_i)
\propto
\lambda_i^{-1},
\qquad
\tilde{\sigma}_i^2
\propto
\lambda_i^{-2}
\end{equation}
The first is the equal-time variance of the fast variable. The second is the noise intensity experienced by the slow variable and contains an additional factor of the fast correlation time,
\begin{equation}
\tau_{\mathrm{corr},i}
=
\lambda_i^{-1} = 1/(1-\beta J_{ii})
\end{equation}
Thus,
\begin{equation}
\tilde{\sigma}_i^2
=
2\nu^2
\operatorname{Var}(x_i)
\tau_{\mathrm{corr},i}
\end{equation}
For a hierarchical network, the corresponding noise intensity may be propagated recursively:
\begin{equation}
\boxed{
s_i
\simeq
\frac{1}{\lambda_i^2}
\left[
\sigma_i^2
+
\sum_{j\neq i}
A_{ij}^2s_j
\right],
\qquad
\tilde{\sigma}_i^2
=
\nu^2s_i
}
\label{eq:RecursiveSlowNoise}
\end{equation}
For a general linearized network, the fluctuation dynamics can be written as
\begin{equation}
d\boldsymbol{\delta x}
=
-K\boldsymbol{\delta x}dt
+
\Sigma d\boldsymbol W
\end{equation}
where $K_{ii}=\lambda_i$, $K_{ij}=-\beta J_{ij}$ for $i\neq j$, and $\Sigma=\operatorname{diag}(\sigma_i)$. The zero-frequency spectral-density matrix is
\begin{equation}
\label{eq:GeneralSlowNoise}
S_x(0)
=
K^{-1}\Sigma\Sigma^{\mathsf T}K^{-\mathsf T}
\end{equation}
The effective slow-noise covariance is
\begin{equation}
D_\theta
=
\nu^2S_x(0)
\end{equation}
This result includes feedback and correlations between different regulatory pathways. Equation~(\ref{eq:RecursiveSlowNoise}) is the corresponding feedforward approximation when correlations between distinct upstream branches can be neglected. A deeper analysis of noise propagation and amplification in networks is given by \cite{rahman2026feedback}.\\


\noindent\textit{Bias-dependent differentiation probability.—}
We determine the probability of differentiation to $\theta_i=\pm1$ for various parameter values. To do so, we use the backward Kolmogorov equation, integrating backward from $\theta_i=1$ and estimating the probability that the system was previously at $\theta_i$. Neglecting the cubic term in Eq.~(\ref{eq:CubicTheta}) gives:
\begin{gather}
    d\theta=a(b+\Lambda\theta)dt + \tilde\sigma dW_t\\
    a=\frac{\nu}{1-\beta J_{ii}}\\
    b = \beta c_i\\
    \Lambda = \beta-1+\beta J_{ii}
\end{gather}
This can then be simplified to:
\begin{equation}
    d\theta_t = \Gamma(\theta_t)dt + \tilde{\sigma} dW_t
\end{equation}
For which we then wish to determine the probability of differentiating to one side:
\begin{equation}
    \mathbb{P}(\theta_0) = \mathbb{P}\{\text{hit $\theta=+1$ before $\theta=-1$ $\vert$ $\theta(0)=\theta_0$}\}
\end{equation}
Which we can determine for the final equilibrium state by solving the backwards Kolmogorov equation:
\begin{equation}
    \Gamma(\theta)u'(\theta) + \frac{\tilde{\sigma}^2u''(\theta)}{2} = 0, \textbf{   }\textit{ }   u(-1) = 0, u(1) = 1
\end{equation}
Asserting $a\sigma\ne 0$, we rewrite and solve for $v(\theta) = u'(\theta)$
\begin{gather}
    v'(\theta) + \frac{2(b+\Lambda\theta)}{a\sigma^2}v(\theta) = 0\\
    u'(\theta) = v(\theta) = C\cdot \exp\left(-\frac{2b\theta + \Lambda\theta^2}{a\sigma^2}\right)
\end{gather}
Integrating this then gives us:
\begin{equation}
    u(\theta) = \int_{\theta_L}^\theta C\cdot \exp\left(-\frac{2b\tau + \Lambda\tau^2}{a\sigma^2}\right)d\tau + D
\end{equation}
To estimate our coefficients $C$ and $D$, let us set $\theta_L = -1$ such that $D=0$ since $u(-1) = 0$. We can estimate $C$ from $u(1) = 1$ such that:
\begin{equation}
    C = \frac{1}{\int_{-1}^1 \exp\left(-\frac{2b\tau + \Lambda\tau^2}{a\sigma^2}\right)d\tau}
\end{equation}
This then gives the differentiation probability:
\begin{equation}
    \mathbb{P}(\theta_0) = \frac{\int_{-1}^{\theta_0} \exp\left(-\frac{2b\tau + \Lambda\tau^2}{a\sigma^2}\right)d\tau}{\int_{-1}^1 \exp\left(-\frac{2b\tau + \Lambda\tau^2}{a\sigma^2}\right)d\tau}
\end{equation}
Rewriting this into error functions and converting to the original reduced parameters we obtain:

\begin{align}
    &\exp\left(-\frac{2b\tau + \Lambda\tau^2} {a\sigma^2}\right)\nonumber\\
    &= \exp\left(\frac{b^2}{a\sigma^2\Lambda}\right)
    \exp\left(-\frac{\Lambda}{a\sigma^2}\left(\tau+\frac{b}{\Lambda}\right)^2\right)\\
    &\gamma = \frac{\Lambda}{a\sigma^2} = \frac{\mu}{\tilde \sigma^2}, \quad
    \epsilon = \frac{b}{\Lambda} = \tilde c
\end{align}
\begin{equation}
\boxed{
    \mathbb{P}(\theta_0) =
    \frac{
    \operatorname{erf}\!\left(\sqrt{\mu/\tilde \sigma^2}\left(\theta_0 + \tilde c\right)\right)
    - \operatorname{erf}\!\left(\sqrt{\mu/\tilde \sigma^2}\left(-1 + \tilde c\right)\right)
    }{
    \operatorname{erf}\!\left(\sqrt{\mu/\tilde \sigma^2}\left(1 + \tilde c\right)\right)
    - \operatorname{erf}\!\left(\sqrt{\mu/\tilde \sigma^2}\left(-1 + \tilde c\right)\right)
    }
}
\end{equation}
To illustrate the analytic expression, we plot several solutions of the hitting probability as a function of the bias $c_i$, in the following figures.\\
\\
\onecolumngrid
\vspace{5mm}
\twocolumngrid
\centering

\includegraphics[width=\linewidth]{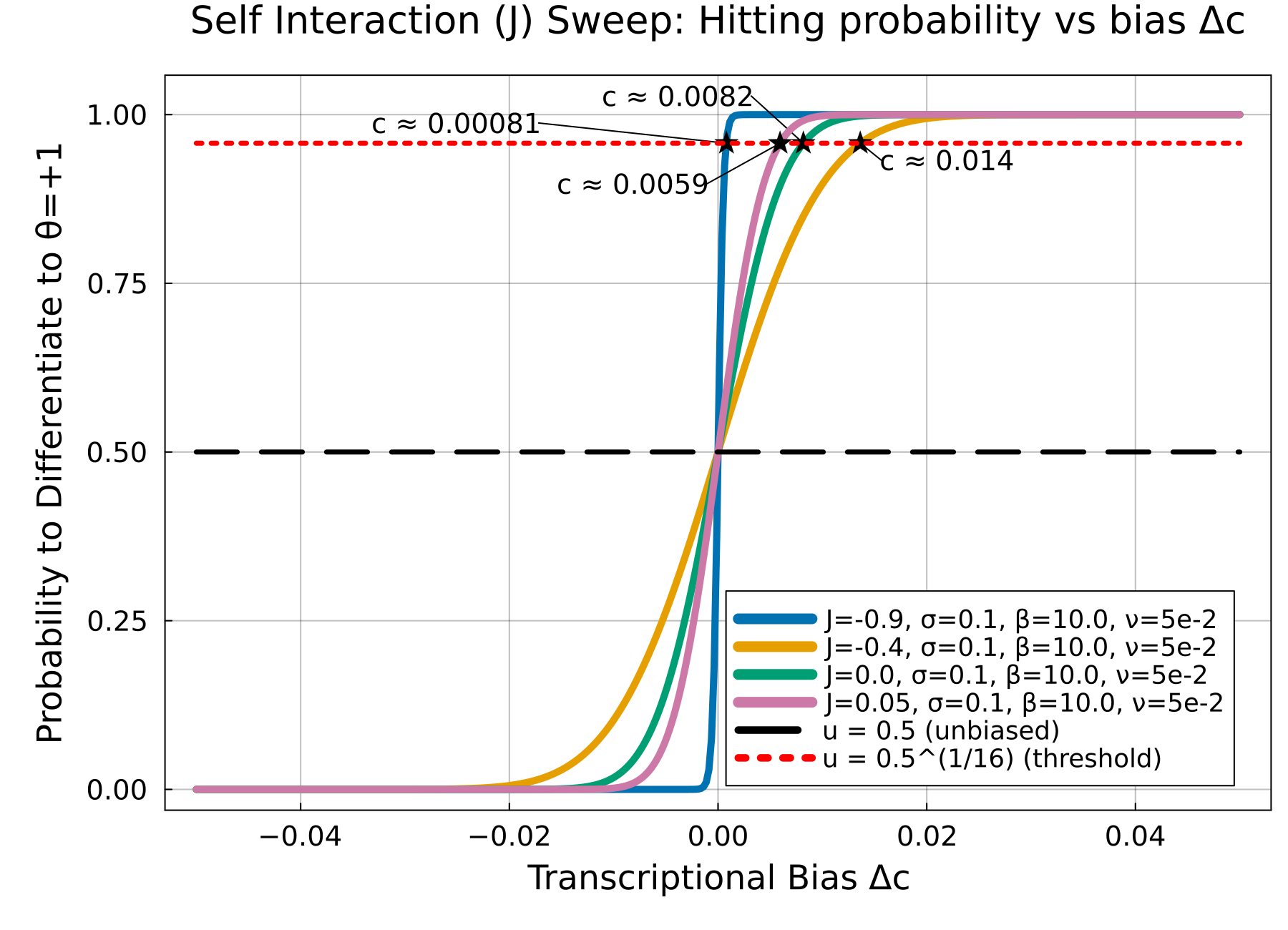}

\vspace{6pt}

\includegraphics[width=\linewidth]{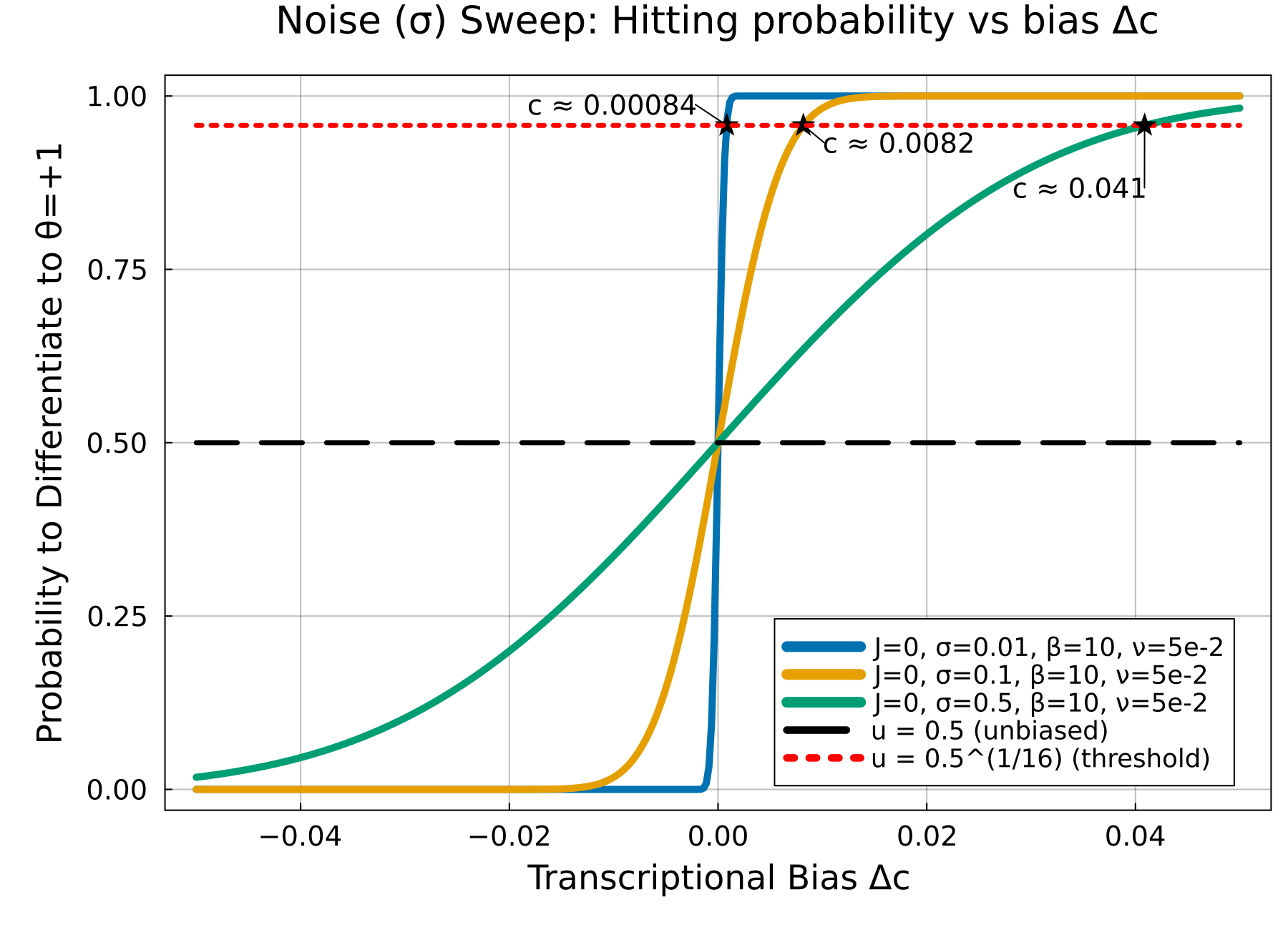}

\vspace{6pt}

\includegraphics[width=\linewidth]{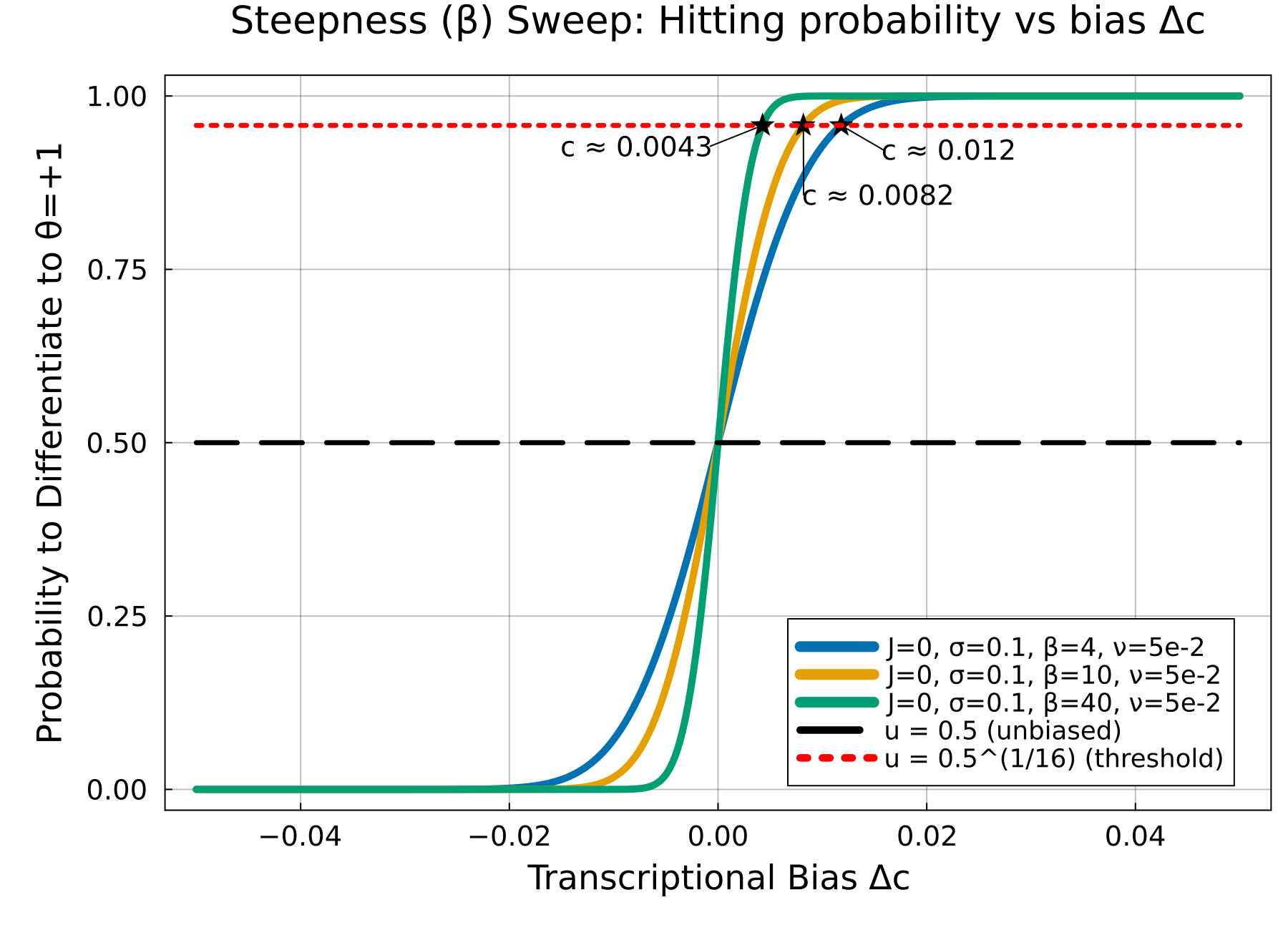}

\vspace{6pt}

\includegraphics[width=\linewidth]{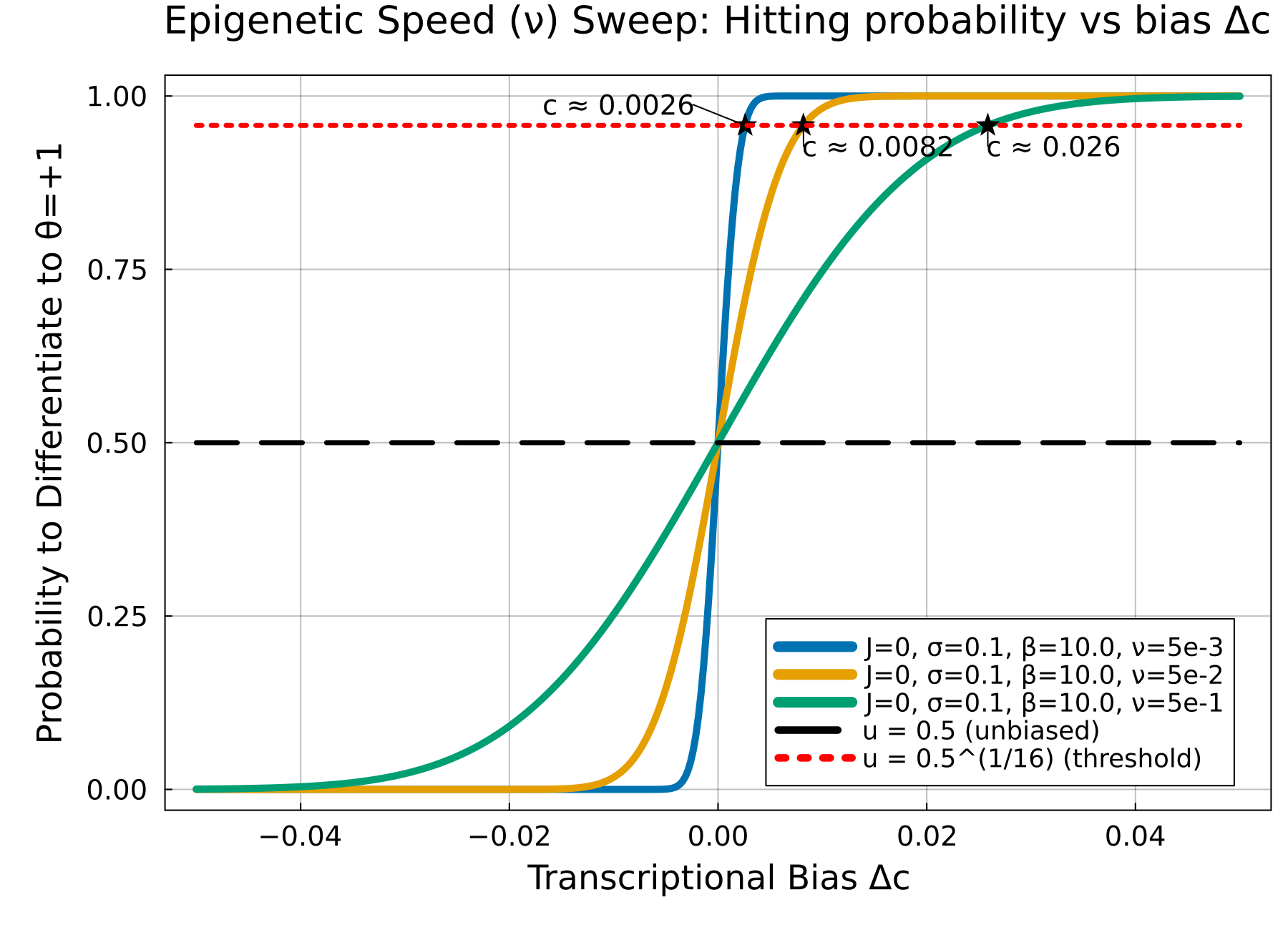}
\onecolumngrid
\captionof{figure}{
Hitting probability as a function of the bias $c_i$ for various values of $J_{ii}$,$\sigma_i$, $\beta$, and $\nu$ respectively. We indicate the intersection at which $P_{0\rightarrow1}(c_i)= \sqrt[16]{1/2}$.
}

\twocolumngrid
\end{document}